\documentclass[12pt]{article}
\begin{document}

\title{\bf \Large Matter Collineations of Some Static Spherically
Symmetric Spacetimes}

\author{M. Sharif \thanks{e-mail: hasharif@yahoo.com}
\\ Department of Mathematics, University of the Punjab,\\ Quaid-e-Azam
Campus Lahore-54590, PAKISTAN.}

\date{}

\maketitle
\begin{abstract}
We derive matter collineations for some static spherically
symmetric spacetimes and compare the results with Killing, Ricci
and Curvature symmetries. We conclude that matter and Ricci
collineations are not, in general, the same.
\end{abstract}


\section{Introduction}

There has been a recent literature [1-11, and references therein]
which shows a significant interest in the study of various
symmetries. These symmetries arise in the exact solutions of
Einstein field equations given by
\begin{equation}
G_{ab}\equiv R_{ab}-\frac 12 Rg_{ab}=\kappa T_{ab},
\end{equation}
where $G_{ab}$ are the components of the Einstein tensor, $R_{ab}$
are the components of Ricci tensor and $T_{ab}$ are the components
of matter (energy-momentum) tensor, $R$ is the Ricci scalar and
$\kappa$ is the gravitational constant.

The well known connection between Killing vectors (KVs) and
constants of the motion [12,13] has encouraged the search for
general relations between collineations and conservation laws
[9,14]. Curvature and the Ricci tensors are the important
quantities which play a significant role in understanding the
geometric structure of spacetime. A basic study of curvature
collineations (CCs) and Ricci collineations (RCs) has been carried
out by Katzin, et al [15] and a complete classification of CCs and
RCs has been obtained by Qadir, et al [4-7,9].

In this paper we shall analyze the properties of a vector field
along which the Lie derivative of the energy-momentum tensor
vanishes. Such a vector is called matter collineation. Since the
energy-momentum tensor represents the matter part of the Einstein
field equations and gives the matter field symmetries. Thus the
study of matter collineations (MCs) seems more relevant from the
physical point of view.

The motivation for studying MCs can be discussed as follows. When
we find exact solutions to the Einstein's field equations, one of
the simplifications we use is the assumption of certain symmetries
of the spacetime metric. These symmetry assumptions are expressed
in terms of isometries expressed by the spacetimes, also called
Killing vectors which give rise to conservation laws [16,17].
Symmetries of the energy-momentum tensor provide conservation laws
on matter fields. These symmetries are called matter
collineations. These enable us to know how the physical fields,
occupying in certain region of spacetimes, reflect the symmetries
of the metric [18]. In other words, given the metric tensor of a
spacetime, one can find symmetry for the physical fields
describing the material content of that spacetime. There is also a
purely mathematical interest of studying the symmetry properties
of a given geometrical object, namely the Einstien tensor, which
arises quite naturally in the theory of General Relativity. Since
it is related, via Einstein field equations, to the material
content of the spacetime, it has an important role in this theory.

There is a growing interest in the study of MCs [10,19,20 and
references therein]. Carot, et al [19] has discussed MCs from the
point of view of the Lie algebra of vector fields generating them
and, in particular, he discussed spacetimes with a degenerate
$T_{ab}$. Hall, et al [20], in the discussion of RC and MC, have
argued that the symmetries of the energy-momentum tensor may also
provide some extra understanding of the the subject which has not
been provided by Killing vectors, Ricci and Curvature
collineations. Keeping this point in mind we address the problem
of calculating MCs for some static spherically symmetric
spacetimes using some special techniques in this paper. It is
hoped that some particular methods can be developed to solve
partial differential equations involving in matter collineation
equations for general spacetimes. This would enable one to obtain
a complete classification of general spacetimes according to their
MCs.

The distribution of the paper follows. In the next section, we
give some basic definitions and write down the matter collineation
equations. In section three we calculate MCs by solving matter
collineation equations for some particular spacetimes. Final
section carries a discussion of the results obtained.

\section{Some Basic Facts and Matter Collineation Equations}

Let $(M,g)$ be a spacetime, $M$ being a Hausdorff, simply
connected, four dimensional manifold, and $g$ a Lorentz metric
with signature (+,-,-,-).

A vector $\xi$ is called a MC if the Lie derivative of the
energy-momentum tensor along that vector is zero. That is,
\begin{equation}
{\cal L}_\xi T=o,
\end{equation}
where $T$ is the energy-momentum tensor and ${\cal L_\xi}$ denotes
the Lie derivative along $\xi$ of the energy-momentum tensor $T$.
This equation, in a torsion-free space in a coordinate basis,
reduces to a simple partial differential equation,
\begin{equation}
T_{ab,c}\xi^c+T_{ac}\xi^c_{,b}+T_{bc}\xi^c_{,a}=0,\quad a,b,c=0,1,2,3.
\end{equation}
where $,$ denotes partial derivative with respect to the
respective coordinate. These are ten coupled partial differential
equations for four unknown functions $(\xi^a)$ which are functions
of all spacetime coordinates.

Collineations can be proper or improper. A collineation of a given
type is said to be {\it proper} if it does not belong to any of
the subtypes. When we solve matter collineation equations,
solutions representing proper collineations can be found. However,
in order to be related to a particular conservation law, and its
corresponding constants of the motion, the {\it properness} of the
collineation type must be known.

We know that every KV is an MC, but the converse is not always
true. As given by Carot et al. [19], if $T_{ab}$ is
non-degenerate, $det(T_{ab})\neq 0$, the Lie algebra of the MCs is
finite dimensional. If $T_{ab}$ is degenerate, i.e.,
$det(T_{ab})=0$, we cannot guarantee the finite dimensionality of
the MCs.

\section{Solution of the Matter Collineation Equations}

We shall solve the MC equations for Minkowski,
Einstein/anti-Einstein, de Sitter/anti-de Sitter, Schwarzschild,
Riessner-Nordstrom and some Bertotti-Robinson like metrics.
However, for Minkowski and Schwarzschild spacetimes, the Ricci
tensor is zero, which implies the vanishing of the energy-momentum
tensor. Hence the MC Eqs.(3) are satisfied for any arbitrary
values of $\xi^a$. Thus in Minkowski and Schwarzschild spacetimes
every $\xi^a$ is a MC.

In de Sitter/anti-de Sitter spacetimes, the energy-momentum tensor
is related to the metric tensor by
\begin{equation}
T_{ab}=\frac {3}{\kappa D^2}g_{ab},
\end{equation}
where $D$ is a constant. If we take the Lie derivative on both
sides of this equation and assume the vanishing of Lie derivative,
the symmetries of the energy-momentum tensor turns out to be
identical to that of the metric tensor. Thus the MCs for these
spacetimes are equal to that of KVs and these are ten.

For Reissner-Nordstrom spacetime, the energy-momentum tensor is
proportional to the Ricci tensor. Thus the MCs become the same as
the four RCs.

We explicitly solve the MC equations for the Einstein metric. However,
we would suffice to give the results for Bertotti-Robinson like
metrics as the same procedure will be applicable for solving MC
equations of these metrics.

For the Einstein metric
\begin{equation}
ds^2=dt^2-\frac{1}{1-r^2/D^2}dr^2-r^2d\Omega^2
\end{equation}
the non-zero components of the energy-momentum tensor are given by
\begin{equation}
T_{00}=\frac {3}{\kappa D^2},\quad T_{ij}=\frac {1}{\kappa
D^2}g_{ij}, \quad (i,j=1,2,3).
\end{equation}
Substituting these values in Eqs.(3), we have
\begin{equation}
\xi^0=A(r,\theta,\phi),
\end{equation}
\begin{equation}
3(1-r^2/D^2)\xi^0_1-\xi^1_{,0}=0,
\end{equation}
\begin{equation}
3\xi^0_2-r^2\xi^2_{,0}=0,
\end{equation}
\begin{equation}
3\xi^0_3-r^2\sin^2\theta\xi^3_{,0}=0,
\end{equation}
\begin{equation}
r\xi^1+D^2(1-r^2/D^2)\xi^1_{,1}=0,
\end{equation}
\begin{equation}
\xi^1+r\xi^2_{,2}=0,
\end{equation}
\begin{equation}
\xi^1+r\cot\theta\xi^2+r\xi^3_{,3}=0,
\end{equation}
\begin{equation}
\xi^1_{,2}+r^2(1-r^2/D^2)\xi^2_{,1}=0,
\end{equation}
\begin{equation}
\xi^1_{,3}+r^2(1-r^2/D^2)\sin^2\theta \xi^3_{,1}=0,
\end{equation}
\begin{equation}
\xi^2_{,3}+\sin^2\theta\xi^3_{,2}=0.
\end{equation}
Solving Eqs.(11)-(16), we have
\begin{equation}
\xi^1=(1-r^2/D^2)^{1/2}[B_3(t)\cos \theta+B_2(t,\phi)\sin\theta],
\end{equation}
\begin{equation}
\xi^2=\frac{(1-r^2/D^2)^{1/2}}{r}[-B_3(t)\sin
\theta+B_2(t,\phi)\cos\theta]+C_1(t,\phi),
\end{equation}
\begin{equation}
\xi^3=\frac{(1-r^2/D^2)^{1/2}}{r\sin\theta}
B_{2\phi}(t)+\cot\theta C_{1\phi}(t,\phi)+E(t,\phi),
\end{equation}
where $B_3$ is a function of time only while $B_2, C_1$ and $E$
are functions of $t$ and $\phi$. Using these equations together
with Eqs.(7)-(10), we finally arrive at the following results
\begin{equation}
\xi^0=\alpha_0,
\end{equation}
\begin{equation}
\xi^1=(1-r^2/D^2)^{1/2}[\alpha_1\cos
\theta+(\alpha_2\cos\phi+\alpha_3\sin\phi)\sin\theta],
\end{equation}
\begin{equation}
\xi^2=\frac{(1-r^2/D^2)^{1/2}}{r}[-\alpha_1\sin \theta+(\alpha_2
\cos
\phi+\alpha_3\sin\phi)\cos\theta]+\alpha_4\cos\phi+\alpha_5\sin\phi,
\end{equation}
\begin{equation}
\xi^3=\frac{(1-r^2/D^2)^{1/2}}{r\sin\theta}(-\alpha_2\sin\phi+\alpha_3\cos\phi)+
\cot\theta(-\alpha_4\sin\phi+\alpha_5\cos\phi)+\alpha_6.
\end{equation}

Thus the MCs are given as
\begin{eqnarray}
\xi=\alpha_0\frac{\partial}{\partial
t}+\alpha_1[(1-r^2/D^2)^{1/2}\cos\theta\frac{\partial}{\partial
r}-\frac{(1-r^2/D^2)^{1/2}}{r}\sin\theta\frac{\partial}{\partial
\theta}]\\\nonumber
+\alpha_2[(1-r^2/D^2)^{1/2}\sin\theta\cos\phi\frac{\partial}{\partial
r}+\frac{(1-r^2/D^2)^{1/2}}{r}\cos\theta\cos\phi\frac{\partial}{\partial
\theta}\\\nonumber
-\frac{(1-r^2/D^2)^{1/2}}{r\sin\theta}\sin\phi\frac{\partial}{\partial\phi}]
+\alpha_3[(1-r^2/D^2)^{1/2}\sin\theta\sin\phi\frac{\partial}{\partial
r}\\\nonumber
+\frac{(1-r^2/D^2)^{1/2}}{r}\cos\theta\sin\phi\frac{\partial}{\partial
\phi}+\frac{(1-r^2/D^2)^{1/2}}{r\sin\theta}\cos\phi\frac{\partial}{\partial
\phi}]\\\nonumber +\alpha_4[\cos\phi\frac{\partial}{\partial
\theta}-\cot\theta\sin\phi\frac{\partial}{\partial
\phi}]\\\nonumber
+\alpha_5[\sin\phi\frac{\partial}{\partial\theta}+\cos\phi\frac{\partial}{\partial
\phi}]+\alpha_6\frac{\partial}{\partial \phi}.
\end{eqnarray}
It follows that the MCs are seven given by
\begin{equation}
\xi_{(1)}=\frac{\partial}{\partial t},
\end{equation}
\begin{equation}
\xi_{(2)}=(1-r^2/D^2)^{1/2}\cos\theta\frac{\partial}{\partial
r}-\frac{(1-r^2/D^2)^{1/2}}{r}\sin\theta\frac{\partial}{\partial
\theta},
\end{equation}
\begin{eqnarray}
\xi_{(3)}=(1-r^2/D^2)^{1/2}\sin\theta\cos\phi\frac{\partial}{\partial
r}+\frac{(1-r^2/D^2)^{1/2}}{r}\cos\theta\cos\phi\frac{\partial}{\partial
\theta}\\\nonumber
-\frac{(1-r^2/D^2)^{1/2}}{r\sin\theta}\sin\phi\frac{\partial}{\partial\phi}],
\end{eqnarray}
\begin{eqnarray}
\xi_{(4)}=(1-r^2/D^2)^{1/2}\sin\theta\sin\phi\frac{\partial}{\partial
r}+\frac{(1-r^2/D^2)^{1/2}}{r}\cos\theta\sin\phi\frac{\partial}{\partial
\phi}\\\nonumber
+\frac{(1-r^2/D^2)^{1/2}}{r\sin\theta}\cos\phi\frac{\partial}{\partial
\phi}],
\end{eqnarray}
\begin{equation}
\xi_{(5)}=\cos\phi\frac{\partial}{\partial\theta}
-\cot\theta\sin\phi\frac{\partial}{\partial\phi},
\end{equation}
\begin{equation}
\xi_{(6)}=\sin\phi\frac{\partial}{\partial\theta}+\cos\phi\frac{\partial}{\partial
\phi}],
\end{equation}
\begin{equation}
\xi_{(7)}=\frac{\partial}{\partial \phi}.
\end{equation}
which are also the generators of a group $G_7$. We see that
$\xi_{(5)},\xi_{(6)},\xi_{(7)}$ are the KVs associated with
spherical symmetry of the Einstein metric. Also, it is to be noted
that MCs $\xi_{(5)},\xi_{(6)},\xi_{(7)}$ are the improper.
Following the same procedure we can evaluate MCs for the
anti-Einstein metric which will be the same as for the Einstein
spacetime except for the difference that the factor $D^2$ will be
replaced by $-D^2$.

There are also three (there may be more) Bertotti-Robinson like
spacetimes available in the literature [21] which are static
spherically symmetric spacetimes. For the first metric given by
\begin{equation}
ds^2_I=(B+r)^2dt^2-dr^2-a^2d\Omega^2,
\end{equation}
where $B$ and $a$ are constants, the only non-zero energy-momentum
components are
\begin{equation}
T_{00}=\frac{(B+r)^2}{\kappa a^2},\quad T_{11}=-\frac{1}{\kappa a^2}.
\end{equation}
Using Eq.(33) in Eq.(3), it follows that
\begin{equation}
\xi^0=-\frac{1}{B+r}(\alpha_0 e^t-\alpha_1 e^{-t})+\alpha_2,
\end{equation}
\begin{equation}
\xi^1=\alpha_0 e^t+\alpha_1 e^{-t},
\end{equation}
\begin{equation}
\xi^2=\xi^2(x^a),
\end{equation}
\begin{equation}
\xi^3=\xi^3(x^a).
\end{equation}
Thus we have
\begin{equation}
\xi=\alpha_0(-\frac{1}{B+r}\frac{\partial}{\partial
t}+\frac{\partial}{\partial
r})e^t+\alpha_1(\frac{1}{B+r}\frac{\partial}{\partial
t}+\frac{\partial}{\partial
r})e^{-t}+\alpha_2\frac{\partial}{\partial
t}+\xi^2(x^a)\frac{\partial}{\partial\theta}+\xi^3(x^a)\frac{\partial}{\partial
\phi}.
\end{equation}
The other two Bertotti-Robinson like metrics are given by
\begin{equation}
ds^2_{II}=\cos^2(c+\sqrt\alpha r)dt^2-dr^2-a^2d\Omega^2,
\end{equation}
\begin{equation}
ds^2_{III}=\cosh^2(c+\sqrt\alpha r)dt^2-dr^2-a^2d\Omega^2,
\end{equation}
where $c, \alpha$ and $a$ are constants. All the diagonal
energy-momentum tensor components survive for these metrics. The
MCs in both cases turn out to be the same as the six KVs for these
metrics.

\section{Conclusion}

We have evaluated MCs for some specific static spherically
symmetric spacetimes. The motivation behind this is to understand
the distribution of matter symmetries as compared to the
symmetries of the metric, Ricci and Curvature tensors [22]. We
discuss the results obtained using the table given above:

\begin{table}
\centering \caption{\bf Comparison of KVs, RCs, CCs, and MCs for
some specific Spherically Symmetric Spacetimes}
\label{Table}
\end{table}

\begin{table}
\begin{tabular}{ccccc}
Metric & KVs & RCs & CCs & MCs \\ \hline Minkowski & 10 &
Arbitrary & Arbitrary & Arbitrary\\ De Sitter/anti & 10 & 10 & 10
&  10 \\ Einstein/anti & 7 & $6+\xi^0(x^a)$ & $6+\xi^0(t)$ & 7\\
Bertotti-Robinson$_{I}$ & 6 & $3+\xi^0(x^a),\xi^1(x^a)$ &
$3+\xi^0(t,r),\xi^1(t,r)$ & $3+\xi^2(x^a),\xi^3(x^a)$\\
Bertotti-Robinson$_{II}$ & 6 &  6 &  6 &  6\\
Bertotti-Robinson$_{III}$ & 6 &  6 &  6 & 6\\ Schwarzschild & 4 &
Arbitrary  &  4 & Arbitrary\\ Reissner-Nordstrom & 4 &  4 &  4 &
4\\
\end{tabular}
\end{table}

It is to be noted that the metric tensor is non degenerate whereas
the Ricci, Riemann and energy-momentum tensors are not necessarily
non degenerate. When there is a degeneracy, it is possible to have
arbitrary collineations. Thus KVs will always be definite but
collineations can be indefinite. Further, if the relevant tensor
vanishes, all vectors become collineations as for Minkowski space
where every vector is a MC. Also, for vacuum spacetime every
vector will be an MC as it is Ricci flat (e.g. Schwarzschild
metric). From the table we see that for Einstein/anti-Einstein
metric MCs are seven while RCs are six plus one arbitrary function
of four variable. This shows that MCs and RCs are not, in general,
the same. However, if $R=0$ then MCs and RCs obviously coincide.

In the Bertotti-Robinson like metric given by Eq.(32), the MCs
turn out to be indefinite in $\xi^2$ and $\xi^3$. The second and
third components of the MCs are functions of all the spacetime
coordinates. This result is different from KVs and even different
from RCs where we have indefinite RCs and CCs in the temporal and
radial components. Thus this type of Bertotti-Robinson like metric
provides new information. It has six KVs, three plus arbitrary
temporal and radial (functions of spacetime coordinates) RCs,
three plus arbitrary temporal and radial (functions of t and r
only) CCs and three plus arbitrary second and third (functions of
all spacetimes) MCs. In the other two Bertotti-Robinson like
metrics, we do not get any information as the KVs, RCs, CCs and
MCs are all the same.

It would be interesting to extend this idea of classifying
spacetimes with some minimal symmetry group in terms of their MCs
and compare the general classification with KVs, RCs and CCs. To
this end, one has to consider the simplest spacetimes which is the
class of all spherically symmetric spacetimes. Then one can extend
this by removing the condition of staticity, moving forward to
cylindrical and plane symmetry and then reducing the minimal
symmetry group further. Finally, it would be worthwhile to make
comparison of the results with other classification schemes, e.g.,
Petrov classification. It is hoped that some interesting feature
will come out.

\newpage

\begin{description}
\item {\bf Acknowledgments}
\end{description}

The author would like to thank Prof. Chul H. Lee for his hospitality at
the Department of Physics and Korea Scientific and Engineering
Foundation (KOSEF) for postdoc fellowship at Hanyang University
Seoul, KOREA.

\vspace{2cm}

{\bf \large References}

\begin{description}

\item{[1]} Tsamparlis, Michael and Apostolopouls, Pantelis S.: J. Math. Phys. {\bf
41}(2000)7573.

\item{[2]} Contreras, G. Nunez, L.A. and Percoco, U.: Gen. Rel. and Grav. {\bf
32}(2000)285.

\item{[3]} Camci, U., Yavuz, $\dot{I}$., Baysal, H., Tarhan, $\dot{I}$. and Yilmaz,
$\dot{I}$.: Int. J. of Mod. Phys. D (to appear 2001).

\item{[4]} Bokhari, A.H., Kashif, A.R. and Qadir, Asghar: J. Math. Phys. {\bf
41}(2000)2167.

\item{[5]} Qadir, Asghar, Ziad, M.: Nuovo Cimento {\bf B113}(1998)773.

\item{[6]} Bokhari, A.H., Qadir, Asghar, Ahmad, M.S. and Asghar, M.: J. Math.
Phys. {\bf 38}(1997)3639.

\item{[7]} Bokhari, A.H., Amir, M.J. and Qadir, Asghar J. Math. Phys. {\bf
35}(1994)3005.

\item{[8]} Melfo, A., Nunez, L. Percoco, U. and Villapba: J. Math. Phys. {\bf 33}(1992)2258.

\item{[9]} Bokhari, A.H. and Qadir, Asghar: J. Math. Phys. {\bf 34}(1993)3543.

\item{[10]} Carot, J. and da Costa, J.: {\it Procs. of the 6th Canadian Conf. on General Relativity
and Relativistic Astrophysics}, Fields Inst. Commun. 15, Amer.
Math. Soc. WC Providence, RI(1997)179.

\item{[11]} Carot, J., Numez, L.A. and Percoco, U.: Gen. Rel. and Grav. {\bf 29}(1997)1223.

\item{[12]} Noether, E.: Nachr, Akad. Wiss. Gottingen, II, Math. Phys. {\bf K12}(1918)235.

\item{[13]} Davis, W.R. and Katzin, G.H.: Am. J. Phys. {\bf 30}(1962)750.

\item{[14]} Bokhari, A.H. and Qadir, Asghar: J. Math. Phys. {\bf 31}(1990)1463.

\item{[15]} Katzin, G.H., Levine, J. and Davis, H.R.: J. Math. Phys.{\bf 10}(1969)617.

\item{[16]} Misner, M.T., Thorne, K.S. and Wheeler, J.A.: {\it
Gravitation} (Freemann, San Francisco, 1973).

\item{[17]} Petrov, A.Z.: {\it Einstein Spaces} (Pergamon, Oxford,
1969).

\item{[18]} Coley, A.A. and Tupper, O.J.: J. Math. Phys. {\bf
30}(1989)2616.

\item{[19]} Carot, J., da Costa, J. and Vaz, E.G.L.R.: J. Math. Phys. {\bf 35}(1994)4832.

\item{[20]} Hall, G.S., Roy, I. and Vaz, L.R.: Gen. Rel and Grav. {\bf 28}(1996)299.

\item{[21]} Qadir, Asghar and Ziad, M.: {\it Procs. of the 6th Marcel Grossmann Meeting}
eds. T. Nakamura and H. Sato (World Scientific 1993);

Ziad, M.: Ph.D. Thesis (Quaid-i-Azam University 1991).

\item{[22]} Bokhari, A.H. and Kashif, A.R.: J. Math. Phys. {\bf 37}(1996)3498.

\end{description}

\end{document}